# Multiple Charge density waves and lattice superstructures in thin-layer TmTe$_2$ and TmTe$_3$

Zhilin Xu (徐智临)[1], Zimeng Zeng (曾梓萌)[1], Shuai-hua Ji (季帅华)[1,2], Xinqiang Cai (蔡新强)[1]**, Xi Chen (陈曦)[1,2†]

[1] State Key Laboratory of Low Dimensional Quantum Physics and Department of Physics, Tsinghua University, Beijing 100084, China

[2] Frontier Science Center for Quantum Information, Beijing, China

*Supported by the National Natural Science Foundation of China under Grant No. 11934001 and No. 11874233, and the Science Challenge Project under Grant No. TZ2016004.
**Corresponding author. Email: cxq15@tsinghua.org.cn
†Corresponding author. Email: xc@mail.tsinghua.edu.cn

**Abstract**

We have grown thulium tellurides (TmTe$_2$, TmTe$_3$) thin layers (less than four layers) on graphene/SiC (0001) by molecular beam epitaxy. The charge density waves (CDWs) and lattice superstructures (LSs) are investigated by scanning tunneling microscopy. Clear CDW patterns in real space are observed on surface of metallic TmTe$_3$. Two CDWs are with wave vectors 0.29**c*** and 0.31**a*** respectively. LSs with various periods are unveiled on the surface of TmTe$_2$ and TmTe$_3$. The electronic structures of these films are semiconducting. These results show that superstructures in rare earth tellurides can have two origins, CDWs or LSs.

**Main text**

Rare earth (R) ditelluride (RTe$_2$) and tritelluride (RTe$_3$) are quasi-layered or layered materials with quasi two-dimensional charge density waves (CDWs)[1-3]. They are composed of ReTe double-corrugated slabs located within squared tiled Te layers, as showed in Fig.1(a). The main difference is that there exists single Te plane sandwiched by RTe slabs for RTe$_2$ and double Te planes for RTe$_3$. The CDW is believed to exist in Te layers with wave vector $\mathbf{q}_{CDW}\sim 1/2\mathbf{a^*}$ for RTe$_2$ [4-6] and incommensurate $\mathbf{q}_{CDW}\sim 2/7\mathbf{c^*}$ for RTe$_3$ [7,8]. For heavy rare earth elements (R=Tb~Tm), a second CDW emerges in RTe$_3$ with wave vector perpendicular to the mentioned one [9-11], giving rise to multiple CDWs in these compounds.

When detecting the CDW states in RTe$_2$ and RTe$_3$, especially by transmission electron microscopy (TEM) or x-ray diffraction (XRD), a notorious misleading effect is the lattice superstructures (LSs) [12], which also displays periodical lattice modulation. The key differences between the LSs and CDW lie in that the former one is purely structurally originated. It has been demonstrated that several types of periodical Te defects in RTe$_2$ can give rise to $\sqrt{5}\times\sqrt{5}$ LSs [13-15]. Furthermore, in R$_2$Te$_5$ compounds which is closely related to RTe$_2$ and RTe$_3$, superstructure patterns attributed to CDW can be found by TEM [16] and XRD [17]. Attentions should be paid to that different modulation wave vectors are observed for Sm$_2$Te$_5$ in these experiments [18]. The discrepancy indicates possible existence of LSs due to sample preparations.

Here, we focus on TmTe$_2$ and TmTe$_3$, typical of RTe$_2$ and RTe$_3$. We prepared the thin layers on graphitized SiC (0001) by molecular beam epitaxy (MBE). By scanning tunneling microscopy/spectroscopy (STM/STS), multiple CDWs are revealed to persistence in few layers TmTe$_3$ and various LSs (at least seven types) are found. The diverse LSs alert careful treatment of superstructures patterns observed by TEM or XRD for rare earth tellurides.

We grew TmTe$_2$ and TmTe$_3$ few layers on single or bilayer graphene prepared on Nitrogen-doped SiC(0001) (~0.1Ω·cm). The base pressure of the MBE system is about 3×10$^{-10}$ torr. Before growth, thulium foils (Alfa, 99.9%) and tellurium shots (Alfa, 99.9999%) were introduced into Knudsen cells and degassed. TmTe$_2$ and TmTe$_3$ thin films were prepared by co-evaporating Tm and Te while the substrate was heated. The vacuum was kept better than 3×10$^{-9}$ torr during the growth. The growth rate is about 0.05ML/min in our experiment. The layer number of prepared films is less than four in our work.

After the growth, samples were transferred to a UNISOKU low temperature STM at 4.4 K (or 78 K with specific statement) without breaking the vacuum. Before the STM measurement, the Pr–Ir alloy tips were checked and modified on the surface of Ag (111) islands grown on Si (111). The STS spectra (or dI/dV spectra) were obtained with bias modulation of 5 mV at 913 Hz.

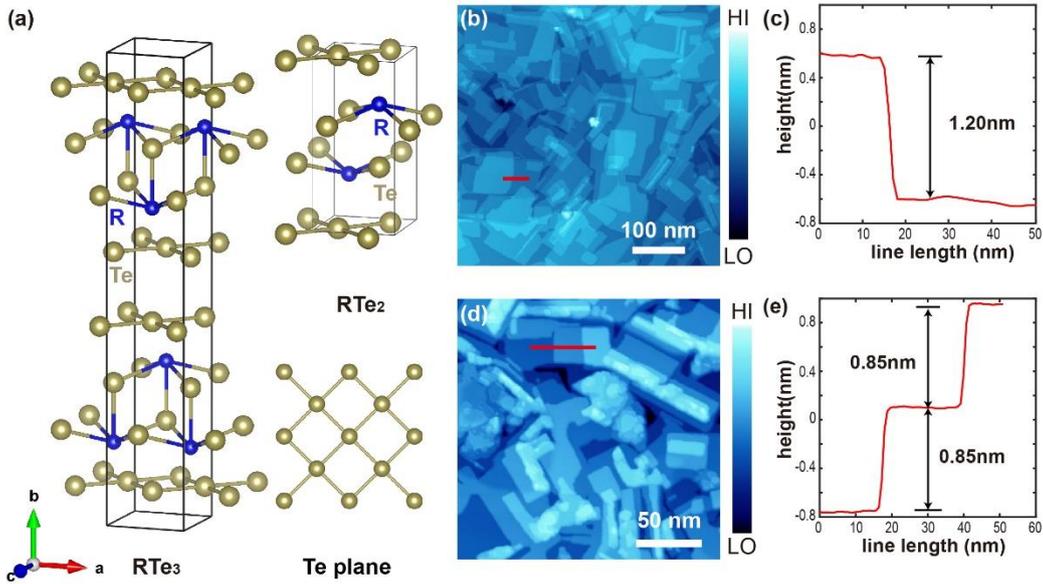

Fig.1 (a) Crystal structure of RTe2 and RTe3. (b) STM topography of layered TmTe$_3$ thin films on graphene (bias: 4 V, current: 20 pA). (c) Step height along the red line indicated in (b). (d) STM topography of layered TmTe$_2$ thin films on graphene (bias: 4 V, current: 20 pA). (e) Step height along the red line indicated in (c).

The topography of prepared thin films is characterized by STM, as showed in Fig. 1. Fig.1 (b) shows the STM topography of a typical sample grown by co-evaporating Tm at 650°C and Te at 265°C for 30 min on a Graphene/SiC substrate kept at 200°C. The apparent heights between layers are about 1.20 nm (see the red line in Fig. 1(c)), corresponding to monolayer (ML) TmTe$_3$ (half of a unit cell of TmTe$_3$). By measuring the heights of steps, we find that most layers correspond to TmTe$_3$. Fig.1 (d) shows the STM topography of a typical sample grown by co-evaporating Tm at 700°C and Te at 260°C for 30 min on a Graphene/SiC substrate kept at 250°C. The apparent heights between layers are about 0.85 nm (see the red line in Fig. 1(e)). They are about the thickness of ML TmTe$_2$ [19]. Clusters with unregulated ingredients can be found in the sample and is not relevant here. Comparison of two samples shows that when the substrate temperature increases from 200°C to 250°C accompanied with decreasing the Te/Tm flux ratio, the main product changes from layered TmTe$_3$ to layered TmTe$_2$.

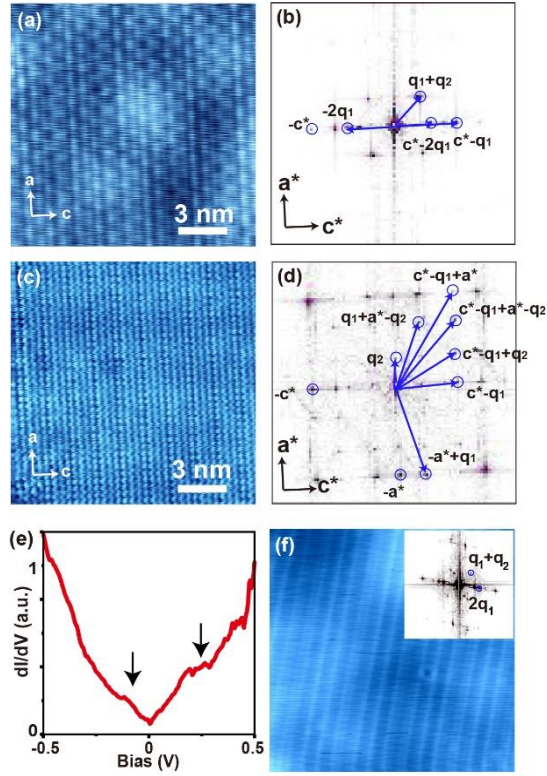

Fig.2 Multiple CDWs of TmTe$_3$. (a) Bias: 1 V, current: 200 pA. (b) FFT image of (a). (c) Bias: 0.3 V, current: 200 pA. (d) FFT image of (c). (e) STS of TmTe$_3$ (set point: 0.5 V, 200 pA). The arrows indicate the CDW gap. (f) Bias: 1 V, current: 100 pA, temperature: 78 K. The inset is the FFT of STM image.

Multiple CDWs can be found on the surface of TmTe$_3$ few layers, as shown in Fig.2. Fig.2(a) and (c) show the STM images of TmTe$_3$ thin film with sample bias set as 1 V and 0.3 V, respectively. The modulations in two directions are clear in Fig. 2(a), showing the existence of two CDW wave vectors. By atomic resolutions in Fig. 2(c), the in-plane lattice constant is measured as 4.28 Å, in agreement with bulk TmTe$_3$ [9]. By fast Fourier transformation (FFT) of STM image (see Fig. 2(b) and (d)), the CDW wave vectors can be resolved. Besides the reciprocal vectors of lattice (**a\*** or **c\***), the dots in FFT image represent the wave vectors of CDWs and their linear superpositions. By comparing with the CDWs of bulk RTe$_3$ (R=Tb~Tm) [9], we can pin down two CDW wave vectors. By measuring these wave vectors on the basis of **a\*** and **c\***, the two CDW wave vectors are found to be **q$_1$**=(0.29 ± 0.01)**c\*** and **q$_2$**=(0.31 ± 0.01)**a\***, perpendicular to each other. Some linear superpositions are also showed in Fig. 2(b) and (d). These results show the persistence of the multiple CDWs in TmTe$_3$ near 2D limit. From the STS spectra of TmTe$_3$ (see Fig. 2(e)), a weak gap feature (indicated by the arrows in Fig. 2(e)) can be seen and local density of state has a residue at Fermi energy, indicating a partially gapped Fermi surface. The multiple CDWs on TmTe$_3$ can still be observed at 78 K (see Fig. 2(f)). But CDWs with **q$_1$** are dominant here, with a stripe pattern in real space. Considering the CDW transition temperature for TmTe3 is T$_1$ = 245 K for **q$_1$** and T$_2$= 185 K for **q$_2$** in bulk TmTe$_3$ [9], the CDW intensity at 78 K should be as 0.7~0.9 times as at 4.4 K. This estimation contracts with our experimental result. The possible reason is lower transition temperature T$_2$ for few layer TmTe$_3$.

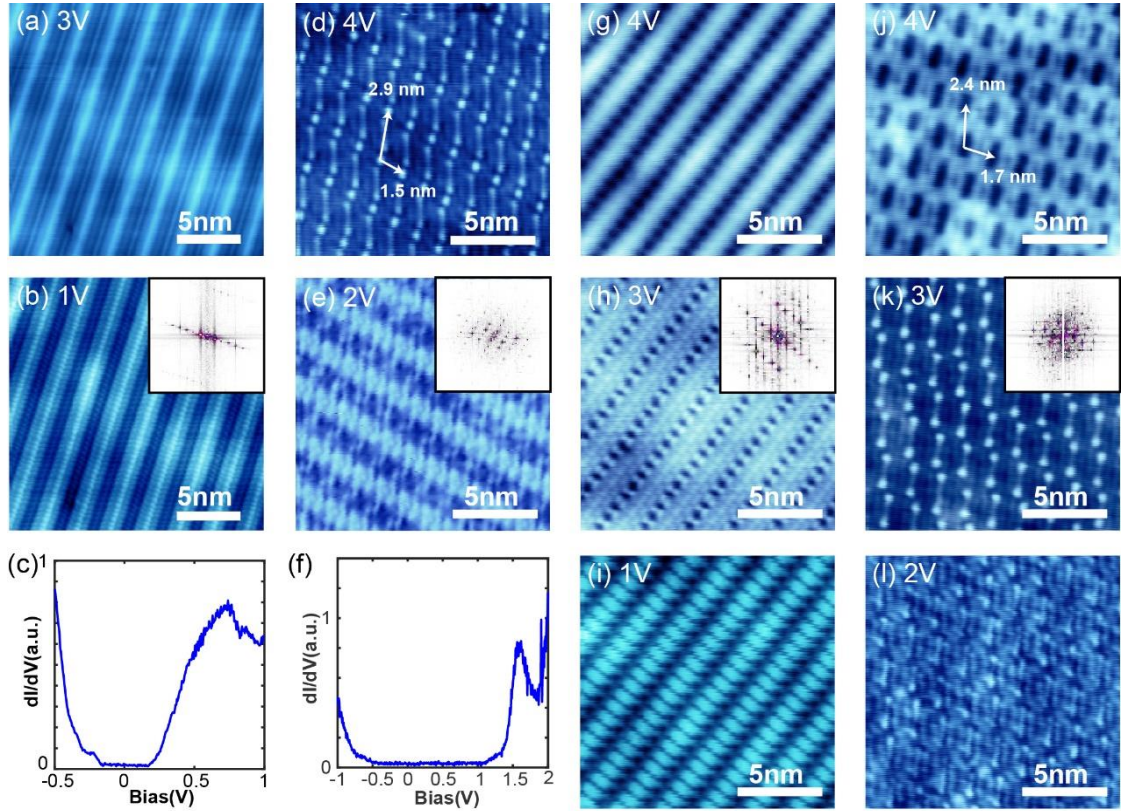

Fig.3 (a) (b) STM topography of LSs type A at 3 V and 1 V. The inset in (b) shows the FFT image. (c) STS of TmTe$_3$ with LSs type A (setpoint: 1 V, 200 pA). (d) (e) STM topography of LSs type B at 4V and 2V. The inset in (e) shows the FFT image. (f) STS of TmTe$_3$ with LSs type B (setpoint: 2 V, 200 pA). (g) (h) (i) STM topography of LSs type C at 4V, 3V and 1V. The inset in (h) shows the FFT image. (j) (k) (l) STM topography of LSs type D at 4V, 3V and 2V. The inset in (k) shows the FFT image.

The LSs are also found on the surface of TmTe$_3$ few layers, as shown in Fig. 3. We observed at least four types of LSs. These patterns are different from the CDW patterns mentioned above. The multiple CDWs disappear in these films. The modulation patterns can still be seen clearly and easily by STM at sample bias as high as 3 V or 4 V. It should be LSs rather than electronic structures driven CDWs. The observed four LSs are termed as A, B, C and D in Fig. 3. Type A is a stripe pattern, as showed in Fig.3(a) and (b). By FFT (see inset of Fig. 3(b)), the wave vector is 0.17**c***. Different from metallic TmTe$_3$ with CDWs, the STS spectra of TmTe$_3$ with LSs type A have an energy gap about 0.5 eV, as showed in Fig.3(c). The wave vectors for type B LSs (see Fig. (d) and (e)) are complex, it's even hard to see the reciprocal lattice vectors clearly in FFT image. This make it hard to get the wave vectors on the basis of reciprocal vectors. In real space, a bone shape patterns can be found (see Fig. 3(d)). The period is about 1.5 nm and 2.9 nm in two directions, as showed in Fig. 3(d). The semiconducting gap is found to be about 1.5 eV (see Fig. 3(f)). The type C LSs (see Fig. (g)~(i)) is clear to be composed of two wave vectors perpendicular to each other, which is 0.21**c*** and 0.48**a*** respectively. The real space pattern is constituted by regular arrangement of rectangle. The type D LSs (see Fig. 3(j)~(l)) is similar to type B, with bone shape patterns (see Fig. 3(k)). The period is about 1.7 nm and 2.4 nm in two directions, as showed in Fig. 3(j).

Various LSs can also be found on the surface of TmTe$_2$. We demonstrate at least three kinds of

LSs, termed as E, F, and G, respectively in Fig. 4. TmTe$_2$ few layers with these LSs are semiconducting. The wave vectors of type E LSs (see Fig. 4(a) and (b)) are 0.25**a**\*±0.17**c**\*. The gap size (see Fig. 4(c)) is found to be about 1.6 eV. The type F LSs (see Fig. 4(d) and (e)) shows a complex pattern. It's like a molecular within a period. The gap is about 2.0 V by STS (see Fig. 4(f)). The type G LSs (see Fig. 4(g) and (h)) is composed of wave vectors 0.5**a**\*, 0.5**c**\*, 0.25**a**\*±0.75**c**\* and 0.75**a**\*±0.25**c**\*. From the STS spectra (see Fig. 4(i)), the semiconducting gap is about 1.0 eV.

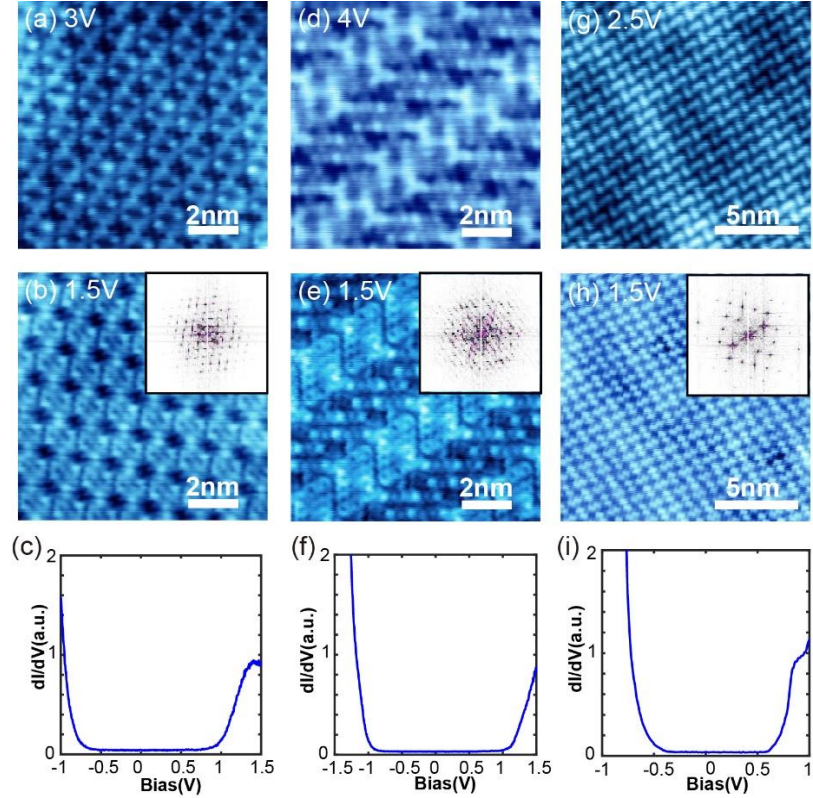

Fig.4 (a) (b) STM topography of surface structure E at 3V and 1.5V. (c) STS of TmTe2 with surface structure E (set point: 1.5 V, 200 pA). (d) (e) STM topography of surface structure F at 4V and 1.5V. (f) STS of TmTe2 with surface structure E (set point: 1.5 V, 200 pA). (g) (h) STM topography of surface structure E at 2.5V and 1.5V, the inset shows the FFT image. (i) STS of TmTe2 with surface structure E (set point: 1 V, 200 pA).

The possible reason for the LSs in TmTe$_2$ and TmTe$_3$ may be the periodical atom deficiency, especially Te deficiency. In RTe$_2$, there exist reports that $\sqrt{5} \times \sqrt{5}$ LSs for LaSe$_{1.90}$ [20], LaTe$_{1.82}$ [12], CeTe$_{1.90}$ [21], SmTe$_{1.80}$ [21] and SmTe$_{1.84}$ [22]. TmTe$_3$ and TmTe$_2$ thin films can become semiconducting when they are not stoichiometric, similar to LaTe$_{1.82}$ [12] and SmTe$_{1.84}$ [22]. Another possible reason of these phenomena is surface reconstructions, where the LSs are just a result of surface effect and do not exist in bulk. These two origins of LSs cannot be distinguished by STM since STM probe cannot reach bulk underneath the surface and further study are needed.

In conclusion, we synthesize few layers of TmTe$_2$ and TmTe$_3$ on Graphene/SiC(0001) by MBE. We find multiple CDWs in TmTe$_3$ thin layer at 4.4 K and 78 K, similar to the bulk materials, showing the persistence of such order near 2D limit. Semiconducting TmTe$_3$ and TmTe$_2$ thin films with various LSs are also demonstrated. These LSs may be periodical Te deficiency.